\documentstyle[multicol,prl,aps,amsfonts,bm,epsfig]{revtex}
\newcommand{\bra}[1]{\mbox{$\langle{#1}|$}}
\newcommand{\ket}[1]{\mbox{$|{#1}\rangle$}}

\def\beq{\begin{equation}}
\def\eeq{\end{equation}}
\def\beqa{\begin{eqnarray}}
\def\eeqa{\end{eqnarray}}

\newcounter{saveeqn}
\newcommand{\alpheqn}{\setcounter{saveeqn}{\value{equation}}%
\stepcounter{saveeqn}\setcounter{equation}{0}%
\renewcommand{\theequation}{\mbox{\arabic{saveeqn}\alph{equation}}}}
\newcommand{\reseteqn}{\setcounter{equation}{\value{saveeqn}}%
\renewcommand{\theequation}{\arabic{equation}}}
\def\beql{\alpheqn \beqa}
\def\eeql{\eeqa \reseteqn}

\begin{document}
%\draft
%\flushbottom

\title{
Modified relative entropy of entanglement for multi-party systems
\thanks{Supported by the National Natural Science Foundation of China under Grant No. 69773052}}
\author{An Min WANG$^{1,2}$}
\address{Laboratory of Quantum Communication and Quantum Computing\\
University of Science and Technology of China$^{1}$\\
Department of Modern Physics, University of Science and Technology of China\\
P.O. Box 4, Hefei 230027, People's Republic of China$^{2}$}
%\date{}

\maketitle
%\bigskip

\begin{abstract}
{We present the modified relative entropy of entanglement for multi-party systems by a given relative density matrix which is spanned by a linear combination of the direct products of so-called basis of relative density matrices and reduced density matrices for every party. For three qubit system in a pure state we derive out the explicit and closed expression of our relative density matrix which is a function of the components of the state vector of three qubits. Through computing  the modified relative entropy of entanglement of some important and interesting examples, we display the main behaviors and elementary properties of the modified relative entropy of entanglement and compare it with the generalized entanglement of formation. Moreover, we also propose an assistant, as the upper bound, of the modified relative entropy of entanglement.}

\medskip
{\noindent}PACS: 03.65.Ud  03.67.-a  \vfill
\end{abstract}

%\smallskip

\begin{multicols}{2}

One of the most active subjects of quantum information is quantum entanglement \cite{Bennett1,Plenio1,Shor}. How to quantify it, distill it, distribute it, product it, use it and so on. Some open questions in these aspects still exist. For example, the universal measures of many-particle system and multi-party system entanglement, if one can be formulated, would have many applications \cite{Cirac,Plenio2}. Aiming at this problem, we had presented \cite{My1} the generalization of entanglement of formation (GEF) which is consistent with the regular definition of entanglement of formation (EF) for bi-partite systems \cite{Bennett2,Wootters}. Moreover, for three qubit system we derive out its explicit and closed expression which is a linear combination of the binary entropy functions with various arguments, and these arguments are clearly determined in terms of the components of state vector of three qubits. In this paper, we again focus on it and present the modified relative entropy of entanglement (MRE) for multi-party system \cite{My2}. 

Just is well known, the relative entropy of the state $\rho$ can be thought as a decent measure of entanglement when the relative density matrix $D$ is such one separable state so that the relative entropy describes a minimum  distant among all of the disentangled states and the state $\rho$ \cite{Vedral}. However, one is short of a general algorithm to find such a relative density matrix $D$ and then computability of relative entropy of entanglement (RE) is not good enough for practical applications. In order to overcome this difficulty and avoid the problem that EF seems to be greater than the entanglement of distillation (DE) \cite{Vedral}, we propose MRE with a determined relative density matrix which depends on the pure state decompositions \cite{My2}. The physical intuition to propose MRE is original from organically combining the advantages of EF for the pure states and RE for the mixed states and avoiding their shortcomings as possibly. For bi-party systems, the obvious expression of the relative density matrix for two qubit systems was obtained, the general definition of the relative density matrix for more qubit systems and the constructive algorithm is also given in principle. However, for more than bi-party systems, the difficulty how to find the relative density matrix still exists now. Although this is a common open question almost for all measures of entanglement, this problem should be fixed since the important applications of entanglement for multi-party systems. For this purpose, we assume, in a pure state precondition, that the relative density matrix $R$ in MRE also can be taken in a ``elongation line" from the state $\rho$ to disentangled states $D$, and the relative entropy $S(\rho\|R)$ generated by it is proportional to the relative entropy $S(\rho\|D)$. Our idea is presented in Fig.1.   
\begin{figure}
\begin{center}
\epsfxsize=2.7in
\epsfysize=2.3in
\epsfbox{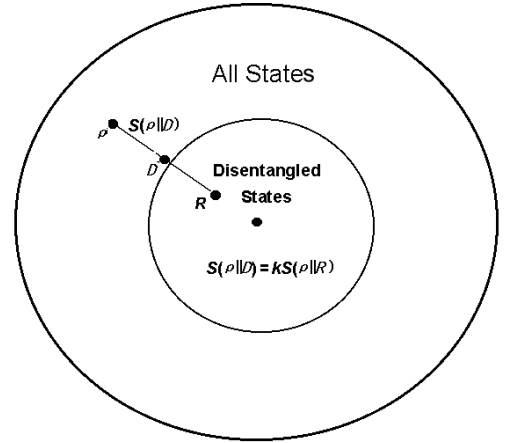}
\vskip 0.1in
\caption{The geometric way to describe the modified relative entropy of entanglement  for multi-party system}
\label{fig1}
\end{center}
\end{figure}
\vskip -0.25in
It must be pointed out that $R$ can not be chosen optionally in the other place and this idea can not be used to a mixed state because the relation between $S(\rho\|D)$ and $S(\rho\|R)$ may be nonlinear, even not exist, and this relation can not be known in general to us. According to this scheme, we present MRE for multi-party systems.  

First, let us recall to MRE in order to introduce our notions. In our paper \cite{My1}, we have obtained the relative density matrix for two qubits in a pure state  
\beq
R_{X\!Y}(\rho_{X\!Y}^{\rm P})=\sum_{j=1}^{2} q^{(j)}(\rho_{X\!Y}^{\rm P}) \bar{\rho}^{(j)}_X(\rho_{X\!Y}^{\rm P})\otimes \bar{\rho}^{(j)}_Y(\rho_{X\!Y}^{\rm P})
\label{2RDM}
\eeq
The coefficients read
\beql
\label{2RDMqa}
q_1(\rho_{X\!Y}^{\rm P})&=&\frac{1-\xi(\rho_{X\!Y}^{\rm P})}{2}\\ 
q_2(\rho_{X\!Y}^{\rm P})&=&1-q_1(\rho_{X\!Y}^{\rm P})
\label{2RDMb}
\eeql
and the basis matrices for $X$- and $Y$- party are 
\beql
\label{2RDMBa}
\bar{\rho}^{(1)}_X(\rho_{X\!Y}^{\rm P})&=&\frac{1}{2}\left[\sigma_0-\bm{\eta}_X(\rho_{X\!Y}^{\rm P})\cdot\bm{\sigma}\right]\\
\bar{\rho}^{(1)}_Y(\rho_{X\!Y}^{\rm P})&=&\frac{1}{2}\left[\sigma_0-\bm{\eta}_Y(\rho_{X\!Y}^{\rm P})\cdot\bm{\sigma}\right]\\
\bar{\rho}^{(2)}_X(\rho_{X\!Y}^{\rm P})&=&\sigma_0-\bar{\rho}^{(1)}_X(\rho_{X\!Y}^{\rm P})\\
\bar{\rho}^{(2)}_Y(\rho_{X\!Y}^{\rm P})&=&\sigma_0-\bar{\rho}^{(1)}_Y(\rho_{X\!Y}^{\rm P})
\label{2RDMd}
\eeql
where $\sigma_0$ is $2\times 2$ identity matrix and $\sigma_k\ (k=1,2,3)$ are usual Pauli Matrices and $\bm{\eta}_X$ and $\bm{\eta}_Y$ are defined by
\beql
\bm{\eta}_{X,Y}(\rho_{X\!Y}^{\rm P})&=&\frac{\bm{\xi}_{X,Y}(\rho_{X\!Y}^{\rm P})}{\xi(\rho_{X\!Y}^{\rm P})}\quad ({\xi(\rho_{X\!Y}^{\rm P})}\neq 0\\
\bm{\eta}_X(\rho_{X\!Y}^{\rm P})&=&\pm\bm{\eta}_Y(\rho_{X\!Y}^{\rm P})=
\{0,0,1\}\quad ({\xi(\rho_{X\!Y}^{\rm P})}=0)\label{eta0}
\eeql
where $\bm{\xi}_X$ and $\bm{\xi}_Y$ are the polarized vectors of reduced density matrices respectively for $\rho_X$ and $\rho_Y$, $\xi_{X\!Y}$ is their norm. In Eq.(\ref{eta0}), the sign is taken as ``$+$" if $\rho_{X\!Y}^{\rm P}=(\ket{00}\bra{00}\pm\ket{11}\bra{11})/\sqrt{2}$, and taken as ``$-$" if $\rho_{X\!Y}^{\rm P}=(\ket{01}\bra{01}\pm\ket{10}\bra{10})/\sqrt{2}$.  We called $\bar{\rho}^{(j)}_X(\rho_{X\!Y}^{\rm P}), \bar{\rho}^{(j)}_Y(\rho_{X\!Y}^{\rm P})$ as the basis of the relative density matrix in a pure state $\rho_{X\!Y}^{\rm P}$ respectively for $X$-party and $Y$-party. If $\rho_{X\!Y}$ is a mixed state with the pure state decompositions $\rho_{X\!Y}^{\rm M}=\sum_i p_{X\!Y}^{(i)}\rho_{X\!Y}^{(i)}$, then
\beqa 
E_{MRE}(\rho_{X\!Y}^{\rm M})&=&\min_{\{p^{(i)},\rho^{(i)}\}\in{\cal{D}}} S\left(\rho_{X\!Y}^{\rm M}||\sum_{i}p_iR(\rho_{X\!Y}^{(i)})\right)\label{IREforM}\nonumber\\
&=&\min_{\{p^{(i)},\rho^{(i)}\}\in{\cal{D}}} S\left(\rho_{X\!Y}^{\rm M}||R^{\rm M}\right)\label{2IRERMM}
\eeqa
where $R(\rho_{X\!Y}^{(i)})$ is a relative density matrix corresponding to the pure state $\rho_{X\!Y}^{(i)}$. It must be pointed out that we can not define directly the basis of relative density matrix in terms of the polarized vectors of reduced density matrices for a mixed state, but we have to find its pure state decomposition corresponding to the minimum values of relative entropy and construct the relative density matrix for a mixed state based on Eq. (\ref{2IRERMM}).    

Now, we can consider the case for the tri-party system. Based on the idea stated above, we present 

\noindent{\bf Definition 1:} For a tri-party system in a pure state, MRE is defined by
\beq
E_{MRE}(\rho_{A\!B\!C}^{\rm P})= \frac{1}{2} S(\rho_{A\!B\!C}^{\rm P}\|R_{A\!B\!C})
\eeq
where the proportional factor $1/2$ comes from the need that the maximum value of MRE is 1 for the three qubit systems and the relative density matrix $R$ is given by the following theorem:  

\noindent{\bf Theorem 1:} In the case of the pure state $\rho_{A\!B\!C}^{\rm P}$ of three qubits
\beqa
\ket{\psi_{A\!B\!C}}&=&a\ket{000}+b\ket{001}+c\ket{010}+d\ket{011}\nonumber\\
& &+e\ket{100}+f\ket{101}+g\ket{110}+h\ket{111}
\eeqa
the relative density matrix $R$ in MRE can be taken as
\beqa
\!R_{A\!B\!C}&\!=\!&\frac{1}{3}\!\sum_{i,j=1}^{2}\!\left[p_{A\!B}^{(i)} q^{(j)}(\rho_{A\!B}^{(i)})\bar{\rho}^{(j)}_A\!(\rho_{A\!B}^{(i)})\!\otimes\! \bar{\rho}^{(j)}_B\!(\rho_{A\!B}^{(i)})\!\otimes\!\rho_{C} \right.\nonumber\\
& &+p_{A\!C}^{(i)} q^{(j)}(\rho_{A\!C}^{(i)})\bar{\rho}^{(j)}_A(\rho_{A\!C}^{(i)})\otimes \rho_{B}\otimes\bar{\rho}^{(j)}_C(\rho_{A\!C}^{(i)}) \nonumber\\
& &\left.+p_{B\!C}^{(i)} q^{(j)}(\rho_{B\!C}^{(i)})\rho_{A}\!\otimes\! \bar{\rho}^{(j)}_B(\rho_{B\!C}^{(i)})\!\otimes\!\bar{\rho}^{(j)}_C(\rho_{B\!C}^{(i)}) \right]
\eeqa
And, the coefficients in above equation read
\beql
p_{A\!B}^{(1)}&=&a{a^*} + c{c^*} + e{e^*} + g{g^*}\label{2RPDAB1}\\
p_{A\!C}^{(1)}&=&a{a^*} + b{b^*} + e{e^*} + f{f^*}\\
p_{B\!C}^{(1)}&=&a{a^*} + b{b^*} + c{c^*} + d{d^*}\label{2RPDBC1}\\
p_{X\!Y}^{(2)}&=&1-p_{X\!Y}^{(1)}\quad (XY=AB,AC,BC)\label{2RPD2}
\eeql
While $q^{(j)}(\rho_{X\!Y}^{(i)})$ are defined by Eqs.(\ref{2RDMqa}-b), where
\beql
\rho_{A\!B}^{(1)}&=&\frac{1}{p_{A\!B}^{(1)}}\left(a\ket{00}+c\ket{01}+e\ket{10}+g\ket{11}\right)\nonumber\\& &\left(a^*\bra{00}+c^*\bra{01}+e^*\bra{10}+g^*\bra{11}\right)\\
\rho_{A\!B}^{(2)}&=&\frac{1}{p_{A\!B}^{(2)}}\left(b\ket{00}+d\ket{01}+f\ket{10}+h\ket{11}\right)\nonumber\\& &\left(b^*\bra{00}+d^*\bra{01}+f^*\bra{10}+h^*\bra{11}\right)\\
\rho_{A\!C}^{(1)}&=&\frac{1}{p_{A\!C}^{(1)}}\left(a\ket{00}+b\ket{01}+e\ket{10}+f\ket{11}\right)\nonumber\\& &\left(a^*\bra{00}+b^*\bra{01}+e^*\bra{10}+f^*\bra{11}\right)\\
\rho_{A\!C}^{(2)}&=&\frac{1}{p_{A\!C}^{(2)}}\left(c\ket{00}+d\ket{01}+g\ket{10}+h\ket{11}\right)\nonumber\\& &\left(c^*\bra{00}+d^*\bra{01}+g^*\bra{10}+h^*\bra{11}\right)\\
\rho_{B\!C}^{(1)}&=&\frac{1}{p_{B\!C}^{(1)}}\left(a\ket{00}+b\ket{01}+c\ket{10}+d\ket{11}\right)\nonumber\\& &\left(a^*\bra{00}+b^*\bra{01}+c^*\bra{10}+d^*\bra{11}\right)\\
\rho_{B\!C}^{(2)}&=&\frac{1}{p_{B\!C}^{(2)}}\left(e\ket{00}+f\ket{01}+g\ket{10}+h\ket{11}\right)\nonumber\\ & &\left(e^*\bra{00}+f^*\bra{01}+g^*\bra{10}+h^*\bra{11}\right)
\eeql
are the needed pure state decomposition. Obviously
\beql
\xi(\rho_{A\!B}^{(1)})&=&{\sqrt{1 - {4
         |ag-ce|^{2}}/{\left(p_{A\!B}^{(1)}\right)^{2}}}}\label{2RPVAB1}\\
\xi(\rho_{A\!B}^{(2)})&=&{\sqrt{1 - {4
         |bh-df|^{2}}/{\left(p_{A\!B}^{(2)}\right)^{2}}}}\label{2RPVAB2}\\
\xi(\rho_{A\!C}^{(1)})&=&{\sqrt{1 - {4
         |af-be|^{2}}/{\left(p_{A\!C}^{(1)}\right)^{2}}}}\label{2RPVAC1}\\
\xi(\rho_{A\!C}^{(2)})&=&{\sqrt{1 - {4
         |ch-dg|^{2}}/{\left(p_{A\!C}^{(2)}\right)^{2}}}}\label{2RPVAC2}\\
\xi(\rho_{B\!C}^{(1)})&=&{\sqrt{1 - {4
         |ad-bc|^{2}}/{\left(p_{B\!C}^{(1)}\right)^{2}}}}\label{2RPVBC1}\\
\xi(\rho_{B\!C}^{(2)})&=&{\sqrt{1 - {4
         |eh-fg|^{2}}/{\left(p_{B\!C}^{(2)}\right)^{2}}}}\label{2RPVBC2}   
\eeql
Thus, $\rho_X^{(j)}(\rho_{X\!Y}^{(i)})$ and $\rho_Y^{(j)}(\rho_{X\!Y}^{(i)}),  (i=1,2)$ can be constructed in terms of Eq.(\ref{2RDMBa}-d). 
 
This theorem can be proved from the properties of relative entropy. It is easy to obtain that

{\noindent}{\bf Corollary 1:}
\beqa
E_{MRE}(\rho_{A\!B\!C}^{\rm P})&\leq&\frac{1}{6}\left[E_{MRE}(\rho_{A\!B})+E_{MRE}(\rho_{A\!C})+E_{MRE}(\rho_{B\!C})\right.\nonumber\\
& &+S(\rho_{A\!B})+S(\rho_{A\!C})+S(\rho_{B\!C})\nonumber\\
& &\left.+S(\rho_{A})+S(\rho_{B})+S(\rho_{C})
+S(\rho_{A})\right]\\
& &\leq E_{GF}(\rho_{A\!B\!C}^{\rm P})
\eeqa
Therefore, the apparent maximum value of MRE is not larger than one of GEF, that is $3/2$ for the three qubit systems. But, the actual maximum value of MRE is 1. The strictly proof will expect to be obtained in future. In fact, we find this maximum value is saturated by all of four pairs of GHZ ``cat" states $
\ket{\phi^{\rm GHZ}}(\pm)=\{(\ket{000}\pm\ket{111})/\sqrt{2}$,
$(\ket{001}\pm\ket{110})/\sqrt{2}$,$\ket{010}\pm\ket{101})/\sqrt{2}$,
$(\ket{011}\pm\ket{100})/\sqrt{2}\}$.  
For 12 kinds of the extended Bell states $
\ket{\psi^{\rm EB}_{AB}}=\ket{\phi^{\rm B}_{AB}}\otimes\ket{\chi_C}$, $\ket{\psi^{\rm EB}_{AC\;1}}=(\ket{0}_A\ket{\chi_B}\ket{0}_C\pm\ket{1}_A\ket{\chi_B}\ket{1}_C)/\sqrt{2}$,$\ket{\psi^{\rm EB}_{AC\;2}}=(\ket{0}_A\ket{\chi_B}\ket{1}_C\pm\ket{1}_A\ket{\chi_B}\ket{0}_C)/\sqrt{2}$,$\ket{\psi^{\rm EB}_{BC}}=\ket{\chi_A}\otimes\ket{\phi^{\rm B}_{BC}}$, MRE are $(1/2)\log_2 3$. While for all the separable states, it is easy to see that their MRE are zero, that is equal to its minimum values. In general cases, we can see MRE has indeed good behaviors and qualifies  correctly the nature of entanglement. For example, for the GHZ-like state $a\ket{000}+h\ket{111}$, $b\ket{001}+g\ket{110}$, $c\ket{010}+f\ket{101}$ and $d\ket{011}+e\ket{100}$, the improved entanglement entropies are respectively equal to the binary entropy functions $H(a a^*)=H(h h^*)$, $H(b b^*)=H(g g^*)$, $H(c c^*)=H(f f^*)$ and $H(d d^*)=H(e e^*)$. Here 
the definition of the binary entropy function $H(x)$ is 
\beq
H(x)=-x\log_2 x-(1-x)\log_2(1-x)
\eeq
The results are just what we expect. For the state
\beqa
\ket{\psi_{ABC}}&=&\frac{x}{3}\ket{000}+\frac{\sqrt{2-x^2}}{3}\ket{001}+\frac{1}{3}\ket{010}\nonumber\\
& &+\frac{1}{\sqrt{6}}\ket{101}+\frac{1}{\sqrt{6}}\ket{110}+\frac{1}{\sqrt{3}}\ket{111}
\label{MyState}
\eeqa
when $x$ varies from 0 to $\sqrt{2}$, we can compute its MRE which is displayed in the following Fig.1  
\begin{figure}
\begin{center}
\epsfxsize=3.2in
\epsfysize=1.8in
\epsfbox{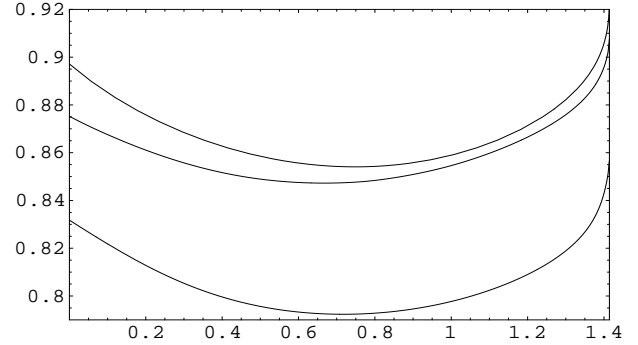}
\caption{The first curve is GEF, the second curve is the assistant of MRE, and the third curve is MRE.}
\end{center}
\end{figure}
\vskip -0.2in

The relations among three kinds of measures of entanglement, GEF, MRE and assistant  of MRE which will be defined a little latter,  are clearly showed in this figure. In special, we can see that MRE for our state (\ref{MyState}) is different an approximate constant from GEF. Perhaps this fact is not an accident and the deep mathematical theorem might hide behind it. This is very interesting open question. Of course, this also implies that our idea is reasonable.

The extension to the mixed state is given in according to the familiar idea \cite{Bennett1,Bennett2}. That is: 
 
\noindent{\bf Definition 2:} For a mixed state $\rho_{A\!B\!C}^{\rm M}$ with the pure state decompositions $\rho_{A\!B\!C}^{\rm M}=\sum_i p_{A\!B\!C}^{(i)}\rho_{A\!B\!C}^{(i)}$, then
\beqa 
E_{MRE}(\rho_{A\!B\!C}^{\rm M})&=&\frac{1}{2}\min_{\{p^{(i)},\rho^{(i)}\}\in{\cal{D}}} S\left(\rho_{A\!B\!C}^{\rm M}||\sum_{i}p^{(i)}_{A\!B\!C}R(\rho_{A\!B\!C}^{(i)})\right)\label{3IREforM}\nonumber\\
&=&\frac{1}{2}\min_{\{p^{(i)},\rho^{(i)}\}\in{\cal{D}}} S\left(\rho_{A\!B\!C}^{\rm M}||R^{\rm M}\right),
\eeqa
where $R(\rho_{A\!B\!C}^{(i)})$ is a relative density matrix corresponding to the pure state $\rho_{A\!B\!C}^{(i)}$. 

We can further suggest the definition of MRE for multi-party systems.

\noindent{\bf Definition 3:} For a n-party system in a pure state, its MRE is defined by
\beq
E_{MRE}(\rho_{A_1\!A_2\!\cdots\!A_n}^{\rm P})=k_nS(\rho_{A_1\!A_2\!\cdots\!A_n} \|R_{A_1\!A_2\!\cdots\!A_n})
\eeq
where the proportional coefficient is
\beq
k_n=\left(\!\sum_{\alpha=1}^{[n/2]}\frac{n!}{\alpha!!(n-2\alpha)!}\!\right)\left(\sum_{\alpha=1}^{[n/2]}\frac{n!(n-\alpha)}{\alpha!!(n-2\alpha)!}\right)^{-1}
\eeq 
and the relative density matrix $R_{A_1\!A_2\!\cdots\!A_n}$ has the form
%\end{multicols}
\beqa
& &R_{A_1\!A_2\!\cdots\!A_n}=\left(\!\sum_{\alpha=1}^{[n/2]}\frac{n!}{\alpha!!(n-2\alpha)!}\!\right)^{-1}\!\sum_{\alpha=1}^{[n/2]}\!\sum_{{s_k,t_k\!=\!1(k=1,\cdots,\alpha)\atop  s_1<s_2<\cdots<s_\alpha}\atop s_k<t_k (k=1,\cdots, \alpha)}^n\nonumber\\ & &\quad\prod_{k=1}^\alpha\left[\sum_{i_{s_k\!t_k}=1}^{n_{s_k\!t_k}} \sum_{j_{s_k\!t_k}=1}^2 p^{(i_{s_k\!t_k})}_{A_{s_k}\!A_{t_k}} q^{(j_{s_k\!t_k})}\left(\rho^{(i_{s_k\!t_k})}_{A_{s_k}\!A_{t_k}}\right)\right]\nonumber\\ & & \quad\left.\begin{array}{c}\\
\displaystyle\bigotimes_{m=1}^n r_{A_m}\\
\\
\end{array}\right|\!\!{}_{{r_{A_{s_k}}=\overline{\rho}^{(j_{s_k\!t_k})}_{A_{s_k}}\big(\rho^{(i_{s_k\!t_k})}_{A_{s_k}\!A_{t_k}}\big),(k=1,\cdots,\alpha)\atop  r_{A_{t_k}}=\overline{\rho}^{(j_{s_k\!t_k})}_{A_{t_k}}\big(\rho^{(i_{s_k\!t_k})}_{A_{s_k}\!A_{t_k}}\big),(k=1,\cdots, \alpha)}\atop r_{A_m}=\rho_{A_m}\;(m\neq s_k,t_k,\;k=1,\cdots,\alpha)} \label{3IRERM}
\eeqa
Here we have assumed that the pure state decomposition of $\rho_{A_{s_k}\!A_{t_k}}$ which leads $S(\rho_{A_{s_k}\!A_{t_k}}|R_{A_{s_k}\!A_{t_k}})$ to be the minimum has the form
\beq
\rho_{A_{s_k}\!A_{t_k}}=\sum_{i_{s_k\!t_k\!}=1}^{n_{s_k\!t_k}} p^{(i_{s_k\!t_k})}_{A_{s_k}\!A_{t_k}} \rho^{(i_{s_k\!t_k})}_{A_{s_k}\!A_{t_k}}
\eeq
While $\rho_{A_m}$ are the reduced density matrices for $A_m$-party and $\bar{\rho}_{A_{s_k}}^{(i_{s_k\!t_k})}$ and $\bar{\rho}_{A_{t_k}}^{(i_{s_k\!t_k})}$ are basis of relative density matrix with respect to $\rho^{(i_{s_k\!t_k})}_{A_{s_k}\!A_{t_k}}$ which is defined in Eq. (\ref{2RDMBa}-d). For bi- and tri-party systems in the pure states, it backs to our standard definitions. In order to understand clearly the physical meaning of MRE, we  introduce 

\noindent{\bf Definition 4:} The assistant of MRE is defined by
\beqa
& &E_{AIR}(\rho_{A_1\!A_2\!\cdots\!A_n}^{\rm P})=\!\left(\sum_{\alpha=1}^{[n/2]}\frac{n!(n-\alpha)}{\alpha!!(n-2\alpha)!}\!\right)^{-1}\sum_{\alpha=1}^{[n/2]}\nonumber\\
& &\left\{\frac{(n-2)!}{(\alpha-1)!!(n-2\alpha)!}\sum_{k_1,k_2=1\atop k_1<k_2}^n  \left[E_{MRE}(\rho_{A_{k_1}\!A_{k_2}})+S(\rho_{A_{k_1}\!A_{k_2}})\right]\right.\nonumber\\& &\left.+\frac{(n-2\alpha)(n-1)!}{\alpha!!(n-2\alpha)!}\sum_{m=1}^n S(\rho_{A_m})\right\}
\eeqa

Obviously, its apparently maximum value is $n k_n$. It is easy to prove that

\noindent{\bf Theorem 2:} 
\beq
E_{MRE}(\rho_{A_1\!A_2\!\cdots\!A_n}^{\rm P})\leq E_{AIR}(\rho_{A_1\!A_2\!\cdots\!A_n}^{\rm P})
\eeq

It can be proved in terms of joint convexity of relative entropy. 
Thus MRE has an apparently maximum value $n k_n$, and its minimum value is 0. 

Again, by use of the familiar idea we have the extension

\noindent{\bf Definition 5:} For a n-party system in a mixed state, its MRE is defined by
\beqa
E_{MRE}(\rho_{A_1\!A_2\!\cdots\!A_n}^{\rm M})
&=&k_n\min_{{\cal{D}}} S(\rho_{A_1\!A_2\!\cdots\!A_n}^{\rm M} \|R_{A_1\!A_2\!\cdots\!A_n}^{\rm M})\\
R_{A_1\!A_2\!\cdots\!A_n}^{\rm M}&=&\sum_i p^{(i)}_{A_1\!A_2\!\cdots\!A_n} R^{(i)}_{A_1\!A_2\!\cdots\!A_n}
\eeqa
where the set ${\cal{D}}$ includes all the possible decompositions of pure states and  $R_{A_1\!A_2\!\cdots\!A_n}(\rho_{A_1\!A_2\!\cdots\!A_n}^{(i)})$ is chosen in according to Eqs.(\ref{3IRERM}). 

We have finished the generalization of MRE for multi-party system in this paper. For three qubit system in a pure state we derive out the explicit and closed expression of our relative density matrix which is a function of the components of the state vector of three qubits. Although it is still very difficulty to find a minimum pure state decomposition for a mixed state, at least for three qubits systems in the pure states, we have obtained the quantitative result of entanglement of multi-party systems. Together with GEF proposed by us \cite{My2}, more knowledge about entanglement will be found in future. 

The one of main advantages of our MRE is that one can finds a determined relative density matrix in terms of a given algorithm (maybe restricted to qubits) and then compute the value of MRE. Moreover, we propose the assistant, as the upper bound, of MRE which is helpful to understand MRE.  However, we pay some prices in two aspects. One is that we have to find the pure state decompositions for a mixed state. It seems to us it is not more difficult than to find a relative density matrix in a huge set of all disentanglement state because the former is to find a decomposition from a given state, but the latter is to find a state from an infinity set of states. The other is that we extend the definition of RE by adding a factor based on our physical idea. From its geometric description Fig.1, our idea is simple and natural. In addition, based the fact that MRE significantly decrease the dependence and sensitivity on the pure state decompositions at least for some interesting states for two qubit system, we also hope that this advantage of MRE can be kept for multi-systems.  

This research is on progressing.

\end{multicols}


\vskip -0.1in 
\begin{references}
\vspace{-0.6in}
\bibitem{Bennett1} C.H.Bennett, H.J.Bernstein, S.Popesu, and B.Schumacher,  {\it Phys. Rev. A} {\bf 53} (1996)2046;  S.Popescu, D.Rohrlich, {\it Phys. Rev. A} {\bf 56} (1997)R3319;
\bibitem{Plenio1} M.B. Plenio and Vedral, {\it Contemp. Phys.} {\bf 39}, 431(1998)
\bibitem{Shor}P. W. Shor.  in {\it Proceedings of the 35th Annual Symposium on the Foundations of Computer Science}, IEEE Computer Society Press, Los Alamitos, CA. ed. S. Goldwasser. 1994. 20; {\it SIAM Journal of Computation} {\bf 26} (1997)1484 
\bibitem{Cirac} W.D\"ur and J.I.Cirac, quant-ph/0011025
\bibitem{Plenio2} M.B. Plenio and V. Vedral, quant-ph/0010080 
\bibitem{My1} An Min Wang, quant-ph/0011040
\bibitem{Bennett2} C.H.Bennett, D.P.DiVincenzo, J.Smolin and W.K.Wootters,  {\it Phys. Rev. A} {\bf 54} (1996)3824
\bibitem{Wootters}W.K.Wootters, {\it Phys. Rev. Lett.} {\bf 80}, 2245(1998); S.Hill and W.K.Wootters, {\it Phys. Rev. Lett.} {\bf 78}, 5022(1997)
\bibitem{My2} An Min Wang, quant-ph/0001023
\bibitem{Vedral}V.Vedral, M.B.Plenio, M.A.Rippin and P.L.Knight,   {\it Phys. Rev. Lett.} {\bf 78} (1997)2275; V.Vedral and M.B.Plenio,  {\it Phys. Rev. A} {\bf 57} (1998)1619
\end{references}
\end{document}